\documentstyle[psfig,aps,prl,amsfonts,amssymb]{revtex}

\begin{document}
\title{Self-Duality, Four-Forms, and the Eight-Dimensional 
Yang-Mills/Dittmann-Bures Field over the Three-Level Quantum Systems} 
\author{Paul B. Slater}
\address{ISBER, University of
California, Santa Barbara, CA 93106-2150\\
e-mail: slater@itp.ucsb.edu,
FAX: (805) 893-7995}

\date{\today}

\draft
\maketitle
\vskip -0.1cm

\begin{abstract}
Utilizing a number of results of Dittmann, we investigate the nature of
the Yang-Mills field  
over the eight-dimensional convex set, endowed with the
Bures metric, of three-level quantum systems. Parallelling 
the decompositions of eight-dimensional
{\it Euclidean} fields by Corrigan, Devchand, 
Fairlie and Nuyts, as well as Figueroa-O'Farrill and others, we
investigate the properties 
of self-dual ($\Omega_{+}$)  and
anti-self-dual  ($\Omega_{-}$)
four-forms  corresponding 
specifically to our {\it Bures/non-Euclidean} context.
For any of a number of 
(nondegenerate) $3 \times 3$ density
matrices, we are able to solve the pair of eigenequations,
$*F = \pm ({1 \over \lambda_{\pm}}) (\Omega_{\pm} \wedge F)$, 
where * is the Hodge operator with respect to the Bures 
metric and $F$ a two-form.
The resultant sets of (traceless) twenty-eight $\lambda_{+}$'s 
coincide with the sets of twenty-eight
$\lambda_{-}$'s, the sets 
consisting of four {\it singlets} and three {\it octets}. The four-forms
$\Omega_{\pm}$ are found to exhibit quite simple behaviors, though we are
not able to derive them in full generality..
\end{abstract}


\vspace{.1cm}

The Bures metric, defined on the nondegenerate density matrices, has been the
object of considerable study 
\cite{uhlmann,hubner1,hubner2,brauncaves,slaterexact}. 
It is the {\it minimal} member of
the nondenumerable family of {\it monotone} metrics \cite{petzsudar,rusles}.
Other members of this family of particular note are the 
``Bogoliubov-Kubo-Mori'' (BKM) 
metric \cite{michor,grasselli}, the maximal monotone metric, as well as
the ``Morozova-Chentsov''  \cite{slatclarke} and 
``quasi-Bures'' \cite{slaterhall} ones, 
the last yielding the minimax/maximin asymptotic
redundancy in universal quantum coding \cite{kratt}.
Interestingly, these (operator) monotone 
metrics correspond in a direct fashion
 to certain
``measures of central tendency'', with, for 
example,  the Bures metric corresponding
to the {\it arithmetic} mean, $(x+y)/2$,  of numbers $x$ and $y$
\cite{petzsudar,rusles}.
All these monotone metrics constitute various extensions to the quantum
domain of the (unique) Fisher information metric in the classical realm
where the objects of study are probability distributions
(rather than density matrices)
\cite{kass} (cf. \cite{frieden}). However,
the only one of these monotone metrics  
 that can be extended to the boundary of pure states
yielding the standard Fubini-Study metric on this boundary
 is the Bures (minimal
monotone) one \cite[sec. IV]{petzsudar} (cf. \cite{bh}).

Dittmann has shown that ``the connection form (gauge field) 
related to the generalization of the Berry phase to the mixed states
proposed by Uhlmann satisfies the source-free Yang-Mills equation 
$*D*D \omega$, where
the Hodge operator is taken with respect to the Bures metric on the space
of finite-dimensional density matrices'' \cite{ditt1} (cf. \cite{rudy}).
``These findings may be seen as extensions to mixed states of numerous 
examples relating the original Berry phase to Dirac monopoles, and the
Wilczek and Zee phase'' \cite{jochenarmin}.
(Let us also indicate here that the BKM metric has been shown to be the
unique monotone Riemannian metric for which two ``natural flat''
connections --- the ``exponential'' and ``mixture'' ones --- are
dual \cite{grasselli}.) 

In  \cite[sec. 6]{ff1},
Figueroa-O'Farrill studied the eigenspaces of various endomorphisms
($\hat{\Upsilon}$),
defined by four-forms ($\Upsilon$) 
in ${\mathbb{E}}^{{8}}$, of the space of two-forms. 
(``One can use $\hat{\Upsilon}$
 to define a generalised {\it self-duality} in eight dimensions by demanding 
that a two-form belong to a definite $G$-submodule of 
$\mathfrak{so_{8}}$.
This generalises self-duality in four dimensions, where we can take
$\Upsilon = *1$, and $\hat{\Upsilon} = *$ itself. The eigenspaces of $\hat{\Upsilon}$ 
are the subspaces of self-dual and anti-self-dual two-forms'' \cite{ff1}.)
We seek to extend this type of  analysis to our different
(non-Euclidean/Bures), but still eight-dimensional
setting.

To proceed, we exploit our recent work 
\cite{slaterjgp1,slaterjgp2} in determining the elements of the
Bures metric using a certain Euler angle parameterization of the
$3 \times 3$ density matrices \cite{byrdslater}. 
(Earlier, Dittmann had noted \cite{ditt1} ``that in affine coordinates
(e. g. using the Pauli matrices for $n = 2$) the [Bures] metric 
becomes very complicated for $n>2$ and no good parameterization seems to
be available for general $n$''.) 
This allows us, among other
things, to employ as our 
parameter space four (noncontiguous)  eight-dimensional hyperrectangles
(each of the four having three sides of length $\pi$, three sides of length 
${\pi \over 2}$,  one  ${\pi \over 4}$ in length and one 
$\cos^{-1}{{1 \over \sqrt{3}}}$), rather than less 
analytically convenient ones, such as that discussed by
Bloore \cite[Fig. 3]{bloore}, involving spheroids and ``tetrapaks''.
(The four hyperrectangles are not adjacent due to the fact that two
of the six Euler angles used in the parameterization have disconnected ranges.
This  situation, that is a multiplicity of 
hyperrectangles  --- which leads to different normalization factors
 only --- has only come to our attention relatively recently, and 
serves as an erratum to earlier analyses \cite{markbyrd1,markbyrd2}, 
which had relied upon work of Marinov \cite{marinov1}, without taking into
account a correction  \cite{marinov2} to \cite{marinov1} (cf. \cite{caves}).)

For the required connection form $A$, Dittmann has presented the 
general formula
\cite[eq. (9)]{ditt1},
\begin{equation} \label{tsd}
A = {1 \over \tilde{L} +\tilde{R}} (W^{*} T - T^{*} W).
\end{equation}
Here the elements of $T$ lie  in the tangent space to the 
principal $U(\mathcal{H})$-bundle, the manifold of invertible normalized 
($\mbox{Tr} W^{*} W = 1$) Hilbert-Schmidt operators, while 
$L$ is the operator (depending on $W$) of left multiplication by
the density matrix 
$\rho = W W^{*}$ and $R$ the corresponding operator of right multiplication.
Also, $\tilde{L}$ and $\tilde{R}$ are the counterpart operators
for $\tilde{\rho} =W^{*} W $.
Since the Euler-angle parameterization of the $3 \times 3$ density 
matrices presented in \cite{byrdslater} is of the (``Schur-Schatten'')
form, 
\begin{equation} \label{Schur}
\rho = U D U^{*}, 
\end{equation}
$U$ being  unitary, $U^{*}$ its conjugate 
transpose, and $D$ 
the diagonal matrix composed 
of the three eigenvalues of $\rho$, one can immediately express $W$ as
$U D^{1/2} U^{*}$, with $W=W^{*}$. Let us note here that 
Sj\"oqvist {\it et al}  have addressed the issue of 
mixed state holonomy --- first raised by Uhlmann 
\cite{uhlmann} as a ``purely mathematical'' 
problem --- developing ``a new formalism of geometric phase for mixed 
states in the experimental context of quantum interferometry'' \cite{sjo}.
In their notation,
\begin{equation}
\Omega = \mbox{i}  \mbox{Tr} [\rho_{0} W^{\dagger} \mbox{d} W]
\end{equation}
is regarded as a gauge potential on the space of density operators, where
\begin{equation}
W(t) = {\mbox{Tr} [\rho_{0} U^{\dagger}(t)] \over
|\mbox{Tr} [\rho_{0} U^{\dagger}(t)]| } U(t),
\end{equation}
the unitary operator $U(t)$ being used to fix the parallel transport
conditions. It would be of interest to examine the relation of 
their expressions
to that of Dittmann (\ref{tsd}), which we implement here 
(cf. \cite{slateruhl}).
(In the context of {\it pure} states, Sanders, de Guise, Bartlett, and Zhang
have presented a scheme
 for  producing and measuring an Abelian geometric
phase shift in a {\it three}-level 
quantum system where states are invariant under a 
non-Abelian group \cite{sanders}. They geodesically evolve
$U(2)$-invariant states in a four-dimensional $SU(3)/U(2)$ space, using
a three-channel interferometer (cf. \cite{byrdgeom}).)

In the non-trivial task (involving computations with modular
operators \cite{rusles}) of implementing formula (\ref{tsd}) 
for the three-level
quantum systems, we relied upon the implicit relation 
(equivalence) between two 
formulas for the Bures metric 
for $n$-level quantum systems \cite[eqs. (2) and (16)]{ditt2},
\begin{equation}
g={1 \over 2} \mbox{Tr} \mbox{d} \rho {1 \over L + R} \mbox{d} \rho,
\end{equation}
and
\begin{equation}
g = {1 \over 2} \sum^{n}_{i,j} a_{ij} \mbox{Tr} \mbox{d} \rho \rho^{i-1} 
\mbox{d} \rho \rho^{j-1}.
\end{equation} (The somewhat involved formula that we used 
for the coefficients
$a_{ij}$, functions of elementary invariants, is stated in Proposition 3
of \cite{ditt2}.)
In other words, we used the formula for the connection
\begin{equation}
A = {1 \over 2} \sum^{3}_{i,j} a_{ij} \rho^{i-1} S \rho^{j-1},
\end{equation}
where $S = W^{*} T - T^{*} W$.

For $n=3$ we have determined, for particular points in the eight-dimensional 
manifold,  four-forms $\Omega_{+}$ for which
$* \Omega_{+} = \Omega_{+}$, where the Hodge star is with respect to the Bures 
metric \cite{slaterjgp1}, as well as 
four-forms $\Omega_{-}$ for which 
$* \Omega_{-} = \Omega_{-}$. We formulated this problem as a set of 
seventy linear simultaneous 
homogeneous equations in seventy unknowns
(that is, unknown {\it constants}). Selecting various points
in our eight-dimensional space 
(and making use of {\it exact} arithmetic), 
MATHEMATICA has consistently succeeded 
in expressing thirty-five of the unknowns 
in terms of the other 
thirty-five variables. However, we did not find a 
a unique such solution valid for arbitrary
points in the eight-dimensional space,
with MATHEMATICA apparently indicating that such a general solution 
over the manifold with
{\it constant} coefficients for $\Omega$
is not possible. 
We have obtained precisely the same type of 
findings in our search for  {\it anti}-self-dual
four-forms. (The Hodge * operation squares to 1 on four-forms in eight 
dimensions. It has eigenvalues +1 and -1 with equal multiplicity. 
So one can always find a basis where * is diagonal with thirty-five 
+1's and thirty-five -1's.) In all these calculations we have used the duality relation,
\cite[eq. (1.5.15)]{azc} 
(making use of our explicit --- though, in some cases somewhat
cumbersome --- formulas for the $g^{ij}$'s \cite{slaterjgp1,slaterjgp2})
\begin{equation} \label{duality}
(* \alpha) _{j_{p+1} \ldots j_{n}} = g^{i_{1} j_{1}} 
\ldots  g^{i_{p} j_{p}} 
\epsilon_{j_{1} \ldots j_{n}} {\sqrt{g} \over p!} \alpha_{i_{1} \ldots i_{p}},
\end{equation}
where we took $n = p = 4$.

For the 
specific point $q_{1}$, we set  the eight variables, 
described in \cite{slaterjgp1}, to be 
$\alpha= - {\pi \over 3}, \tau= {2  \pi \over 3},a={3 \pi \over 
4}, b=-{2 \pi  \over 3},\beta= {2  \pi \over 3},
\theta= -{2 \pi \over 3},\theta_{1} = -{2 \pi \over 3}, \theta_{2} = - {\pi 
\over 3}$ --- the first six 
of these being Euler angles that parameterize the 
unitary matrix $U$, while  the last two parameterize the diagonal matrix
of eigenvalues $D$ in the decomposition (\ref{Schur}). In our 
eight-dimensional manifold of states, the point $q_{1}$ gives us the specific
$3 \times 3$ density matrix,
\begin{equation} \label{denmat}
\rho_{1}  = \pmatrix{ {307 \over 1024} & -{3 (59 \mbox{i} + 25 \sqrt{3}) 
\over 2048} & {(-1)^{5/12} (51 + 29 \mbox{i} \sqrt{3}) \over 1024}
\cr - {3 (- 59 \mbox{i} + 25 \sqrt{3}) \over 2048} & {417 \over 1024} 
& -{3 (-1)^{1/12} (21 \mbox{i} + 25 \sqrt{3}) \over 1024} \cr
{1 \over 512} (-1)^{11/12} (9 + 20 \mbox{i} \sqrt{3}) 
& {3 \over 512} (-1)^{3/4} (24 + \mbox{i} 
\sqrt{3}) & {75 \over 256} \cr}.
\end{equation}
This matrix is, of course, Hermitian, nonnegative 
definite --- possessing eigenvalues $ {9 \over 16}, {1 \over 4}, 
{3 \over 16}$ --- and has trace 1. (It corresponds to a strictly impure or
mixed state. Also, we should point out that the entries of the 
$8 \times 8$ Bures metric tensor
turn out, in fact, to be independent of the variables $\alpha$ and $a$, as
established in \cite{slaterjgp1}.)
For the solutions for $\lambda_{+}$ at $q_{1}$  of the eigenequation, 
($F$ being a two-form),
\begin{equation} \label{kjd}
*F = {1 \over \lambda_{+}} (\Omega_{+}  \wedge F),
\end{equation}
(having substituted unity for  
each of the free thirty-five variables in the 
thirty-five-dimensional solution to the set of seventy
linear homogeneous equations for $\Omega$, 
based upon (\ref{duality}), which  enforce {\it self-duality}),
we obtained the twenty-eight eigenvalues (cf. \cite[eq. (2.1)]{corrigan})
\begin{equation} \label{plmn}
\lambda_{+}(q_{1})  = 6.15149, -6.06045, \pm 5.11128 (\mbox{eightfold}),
-4.16211, 4.07107, \pm .994689 (\mbox{eightfold}),
\pm .0455182  (\mbox{eightfold}).
\end{equation}
Of course, the three octets are each  split into two quartets of opposite 
sign. (The adjoint representation of $SU(3)$ is eight-dimensional, with its
elements generically lying in $SO(8)$ \cite[sec. 7]{byrdsudar}.)
The four singlets or isolated values are the roots of the 
quartic (biquadratic) equation,
\begin{equation} \label{firsteq}
17848517231861271296 + 718875 \lambda_{+} (-72767012864 -2131611323040 
\lambda_{+}
+ 39304490625 \lambda_{+}^3) = 0.
\end{equation}
The six possible values assumed by the three octets are the roots
of the sextic equation (a cubic equation in $x^{2}$),
\begin{equation} \label{sextic}
82734971267961585664 - 18225 \lambda_{+}^{2} (2195802859754043904 +
291144375 \lambda_{+}^2 (-7894856752 + 291144375 \lambda_{+}^{2})) = 0 .
\end{equation}
(Consistent with tracelessness, the twenty-eight eigenvalues sum to 
zero --- as well as, therefore, the four isolated eigenvalues themselves.)
We have (as one particular 
example) ``simplified'' the exact symbolic expression for the root
 of (\ref{sextic}) corresponding to the eigenvalue 5.11128, so that it is 
explicitly real-valued, obtaining thereby that
\begin{equation} \label{term}
(5.11128)^2 \approx {1 \over 873433125} \Big( 16 \Big( 493428547 + 
\end{equation}
\begin{displaymath}
2 \sqrt{217739666231788507} \cos{ \Big[{1 \over 3}  \mbox{tan}^{-1} \Big[
{19986057 \sqrt{257834787813597115559383045701069731} \over
101094855629270248323646732 }} \Big] \Big]  \Big) \Big).
\end{displaymath}
(The value of the cosine term here is close to unity, that is .999444.
Any further simplification of (\ref{term}) does not seem possible.)

We should also note that the cardinality of the set  
of eigenvalues, that is twenty-eight, corresponds
to the number of entries of the
(antisymmetric) two-form $F$, that is $F_{ij}$ ($i, j = 1,...,8)$, 
for which $i<j$ (having observed that $F_{ji}= - F_{ij}$ and $F_{ii} = 0$).
Proceeding  in the exact same manner, but 
using the {\it anti-self-dual} form MATHEMATICA provided, and solving
\begin{equation} \label{kjdd}
*F = - {1 \over \lambda_{-}} (\Omega_{-} \wedge F),
\end{equation}
we obtained precisely the same set of twenty-eight eigenvalues 
(\ref{plmn}), despite
that fact that $\Omega_{+} \neq \pm \Omega_{-}$.

We observe here that the 
well-known 
octonionic instanton (CDFN) equations \cite{corrigan} are obtained by 
choosing a particular constant (hence closed) four-form in
${\mathbb{R}}^{8}$. This form decomposes the bundle of two-forms into two
sub-bundles of dimensions seven and twenty-one. Each of these sub-bundles
corresponds to a different nonzero eigenvalue (-3 and 1, respectively) 
of the transformation (which is symmetric, hence diagonalizable, as well 
as traceless)
\begin{equation}
F \mapsto *(\Omega \wedge F),
\end{equation}
so restricting a Lie-algebra valued two-form to lie in either of
these two sub-bundles guarantees that it will satisfy the Yang-Mills
equations of motion.

We have (denoting by $\zeta_{ijkl}$ 
the four form $\mbox{d} x_{i} \wedge \mbox{d} x_{j}
\wedge \mbox{d} x_{k} \wedge \mbox{d} x_{l}$, where  for 
our eight variables we take $x_{1} =\alpha,
x_{2} = \tau, x_{3} =a, x_{4} =\beta, x_{5} = b, x_{6} = \theta,
x_{7} = \theta_{1}, x_{8} =\theta_{2}$, and using lexicographic ordering 
of the indices), corresponding to the point $q_{1}$,
\begin{equation} \label{fourform1}
\Omega_{+}(q_{1}) = {378375 \over 1654016} \zeta_{1234} -
{14037 \over 127232} \zeta_{1235} \ldots +{601 \over 71} 
\zeta_{1578}
-{59079 \over 284} \zeta_{1678} +  \zeta_{2345} \ldots + \zeta_{5678},
\end{equation}
and 
\begin{equation} \label{fourform2}
\Omega_{-}(q_{1}) = {6975 \over 23296} \zeta_{1234} 
-{14187 \over 127232} \zeta_{1235} \ldots  -{647 \over 71} \zeta_{1578} + 
{58745 \over 284}
\zeta_{1678} +   \zeta_{2345} \ldots + \zeta_{5678} .
\end{equation}
So we see that $\Omega_{+}(q_{1}) \neq \pm \Omega_{-}(q_{1})$ (nevertheless, 
clearly sharing
certain features).
When we used these {\it same} two four-forms, but at another point  
($\alpha= {2 \pi \over 3}, \tau= {2 \pi \over 3},
a= {5 \pi \over 6}, b = {\pi \over 3}, \beta= {\pi \over 4}, \theta= 
{\pi \over 6}, \theta_{1} = {\pi \over 4}, \theta_{2} ={\pi \over 6}$)
than that yielding $\rho_{1}$, 
the two sets of eigenvalues for the equations (\ref{kjd}) 
and (\ref{kjdd}) were quite similar 
(e. g. listing 
the largest ones in absolute value, 9.83657 vs. 9.66359, 
-9.73817 vs. -9.59167,...) but not now 
strictly equal, and the
twenty-eight eigenvalues in each set were  all distinct from one another.

We, then, studied a second density matrix (cf. (\ref{denmat})),
\begin{equation} \label{denmat2}
\rho_{2}  = 
\pmatrix{ {41 \over 128} & -{1 \over 32} +{15 \mbox{i} \over 128} &
(-1)^{11/12} (1+(-1)^{1/3}) \cr
-{1 \over 32} -{15 \mbox{i} \over 128} & {41 \over 128} &
(-{3 \over 128} -{7 \mbox{i} \over 128}) ((-1)^{5/12} +(-1)^{3/4}) \cr
({3 \over 128} +{7 \mbox{i} \over 128}) ((-1)^{1/4} +(-1)^{7/12}) &
 {({1 \over 64} + {5 \mbox{i} \over 128}) (-3 \mbox{i} +\sqrt{3}) \over
\sqrt{2}} & {23 \over 64} \cr},
\end{equation}
having eigenvalues ${1 \over 2}, {3 \over 8}$ and ${1 \over 8}$.
It corresponds to a 
new point $q_{2}$ for which
\begin{equation} \label{q2}
 \alpha = {\pi \over 4}, \tau= {3 \pi \over 4},
a = {2 \pi \over 3}, b= {\pi \over 4}, \beta = {\pi \over 4},
\theta= {\pi \over 3}, \theta_{1} = {\pi \over 4}, \theta_{2} = 
{\pi \over 6}.
\end{equation}
Adopting the same analytical 
strategy as employed for $\rho_{1}$, we obtained 
for the solutions for $\lambda_{+}$ 
at $q_{2}$  of the equation (\ref{kjd}),
the twenty-eight eigenvalues (cf. (\ref{plmn})),
\begin{equation} \label{las}
\lambda_{+}(q_{2}) = 2.68934, -2.60397,  \pm 2.14857 (\mbox{eightfold}), 
-1.69317, 1.6078,
\pm  .498082 (\mbox{eightfold}), .0426838 (\mbox{eightfold}).
\end{equation}

So again, fully analogously to (\ref{plmn}), we have four singlets and
three octets. 
(Tracelessness is again verifiable. Let us also note that for 
the  analysis based on 
$\rho_{1}$, as well as  that on $\rho_{2}$,
the four singlet eigenvalues have the largest, 
second largest, and fourth and fifth largest of the seven possible distinct 
absolute values.) 
Computations of the analogues based on $\rho_{2}$ 
of the {\it exact} results
(\ref{firsteq}), (\ref{sextic}) and (\ref{term}) 
proved now to be much more intractable,
though.
The self-dual four-form here employed was
\begin{equation} \label{s1}
\Omega_{+}(q_{2})  = { 448 + 128 \sqrt{3} + 27 \sqrt{6} \over 896} \zeta_{1234}
\ldots + {-3 +224 \sqrt{6} \over 6} 
\zeta_{1678} + \zeta_{2345} \ldots + \zeta_{5678},
\end{equation}
while the anti-self-dual four-form was
\begin{equation} \label{s2}
\Omega_{-}(q_{2}) = {448 + 128 \sqrt{3} -27 \sqrt{6} \over 896}
 \zeta_{1234} \ldots + {-3 - 224 \sqrt{6} \over 6}
\zeta_{1678} + \zeta_{2345} \ldots + \zeta_{5678}.
\end{equation}
We found the same set of twenty-eight
eigenvalues (\ref{las})  relying upon $\Omega_{-}(q_{2})$, using the
equation (\ref{kjdd}),  as we did on
$\Omega_{+}(q_{2})$, using the equation (\ref{kjd}).

For several additional density matrices (other than $\rho_{1}$ and 
$\rho_{2}$), we have been able to obtain fully analogous results.
In all these instances, we let the ``second'' (lexicographically speaking)
set of thirty-five variables (that is, the coefficients of 
$\zeta_{2345} \ldots
\zeta_{5678}$) be unity, and solved for the coefficients of the ``first''
set of thirty-five ($\zeta_{1234} \ldots \zeta_{1678}$).

We have also been able to derive univariate generalizations of the 
(constant) four-forms
(\ref{s1}) and (\ref{s2}), allowing either the Euler angle $\tau$, $\beta$,
$b$
or $\theta$   to be free, rather than fixed at particular values.
It turns out --- as numerical evidence firmly convinced us --- that 
the 
twenty-eight 
eigenvalues (\ref{las}) are completely unchanged as $\beta$ is varied
(even though the coefficients of the extended four-form 
themselves do explicitly depend on $\beta$),
but do change with $\tau$, $b$ or $\theta$.
For example, for the point $q_{2}$ (\ref{q2}) 
 we have obtained the four-form
(\ref{s1}). The coefficient there of $\zeta_{1234}$ is
\begin{equation} \label{kdoe}
c_{1234} = {448 + 128 \sqrt{3} + 27 \sqrt{6} \over 896} \approx .821249.
\end{equation}
If we let the Euler angle $\tau$ vary 
(its full range being $[0,\pi]$) rather than being
fixed at $3 \pi /4$ (as it is, of course, for $q_{2}$ itself), 
we obtain instead of (\ref{kdoe}) the 
more general expression,
\begin{equation} \label{h1}
c_{1234}(\tau) = 
\end{equation}
\begin{displaymath}
{28 \sqrt{2} (42 - 5 \sqrt{3}) \cos{2 \tau}
+ 119 \sqrt{6} \cos{4 \tau} - 4480 (7 + 2 \sqrt{3}) \sin{2 \tau}
+\sqrt{6}  (2009 -10 \sin{4 \tau}) + 84 \sqrt{2} \sin{ 4 \tau} \over 62720}.
\end{displaymath}
Similarly, letting $\beta, b$  and $\theta$ vary in turn we derive,
\begin{equation} \label{h2}
c_{1234}(\beta) = {1 \over 996} (448 + 128 \sqrt{3} + 27 \sqrt{6})
\sin{2 \beta},
\end{equation}
\begin{equation} \label{h3}
c_{1234}(b) = {1 \over 2} + {\sqrt{3} (128 + 27 \sqrt{2}) \sin{2 b} 
\over 128 (7 + \cos{2 b})},
\end{equation}
and
\begin{equation} \label{h4}
c_{1234}(\theta) = 
\cos{\theta} + {2 \sin{\theta} \over 7} + {9 \cos{\theta} 
\sin^{3}{\theta} \over 28 \sqrt{2}}.
\end{equation}
Also,
\begin{equation} \label{h5}
c_{1234}(\theta_{1})=
\end{equation}
\begin{displaymath}
 { 1600 + 256 ( 7 + 8 \sqrt{3}) \cos{2 \theta_{1}}
+64 (11+14 \sqrt{3}) \cos{4 \theta_{1}} + 3 \sqrt{3} 
(384 + 26 \sin{\theta_{1}} + 35 \sin{3 \theta_{1}} + 25 \sin{5 \theta_{1}})
 \over 128 (25 + 28 \cos{2 \theta_{1}} + 11 \cos{4 \theta_{1}}}
\end{displaymath}
(We have not so far --- due to  increased computational 
burdens --- been able to similarly analyze the scenarios in which 
$\theta_{2}$ is free, nor in which {\it pairs}
of the parameters can simultaneously vary.)
In Figures 1 - 5 we plot these five  simple well-behaved 
univariate functions (\ref{h1})-(\ref{h5}), which generalize the
constant coefficient (\ref{kdoe}) 
over the ranges of their respective variables.
\newpage
\begin{figure}
\centerline{\psfig{figure=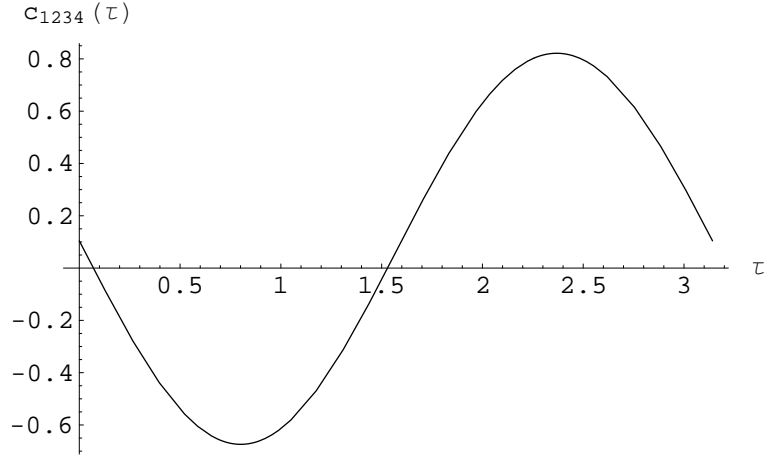}}
\caption{Univariate generalization (\ref{h1}), 
incorporating the Euler angle
$\tau$, of the constant coefficient (\ref{kdoe}) of $\zeta_{1234}$ in 
the self-dual 
four-form $\Omega_{+}(q_{2})$ (\ref{s1}), corresponding to 
the density matrix $\rho_{2}$ 
(\ref{denmat2})}
\label{f1}
\end{figure}
\begin{figure}
\centerline{\psfig{figure=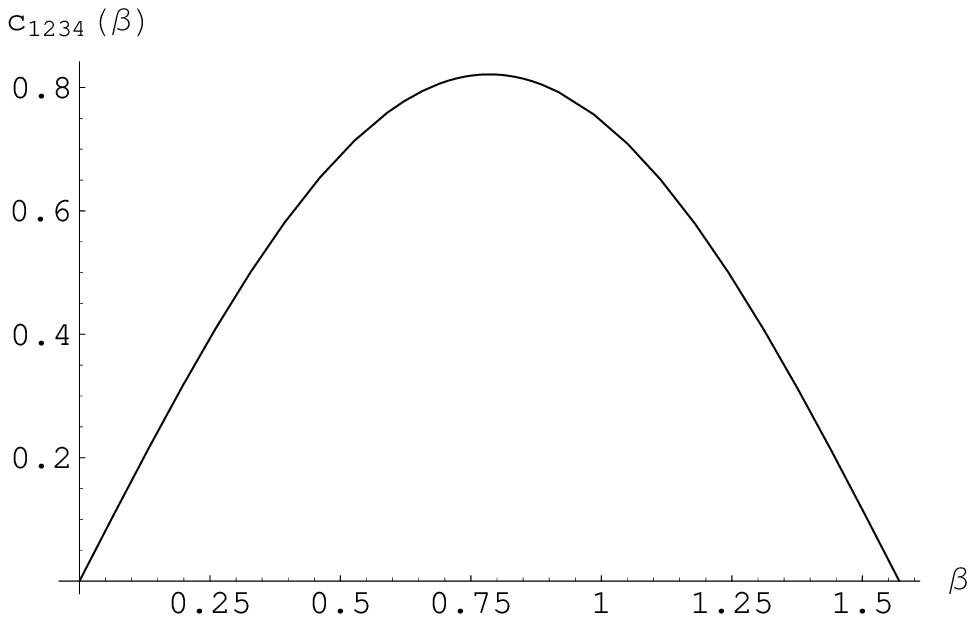}}
\caption{Univariate generalization (\ref{h2}), incorporating the Euler angle
$\beta$, of the constant coefficient (\ref{kdoe}) of $\zeta_{1234}$
in the self-dual 
four-form $\Omega_{+}(q_{2})$ (\ref{s1}), corresponding to 
the density matrix $\rho_{2}$ 
(\ref{denmat2})}
\label{f2}
\end{figure}
\begin{figure}
\centerline{\psfig{figure=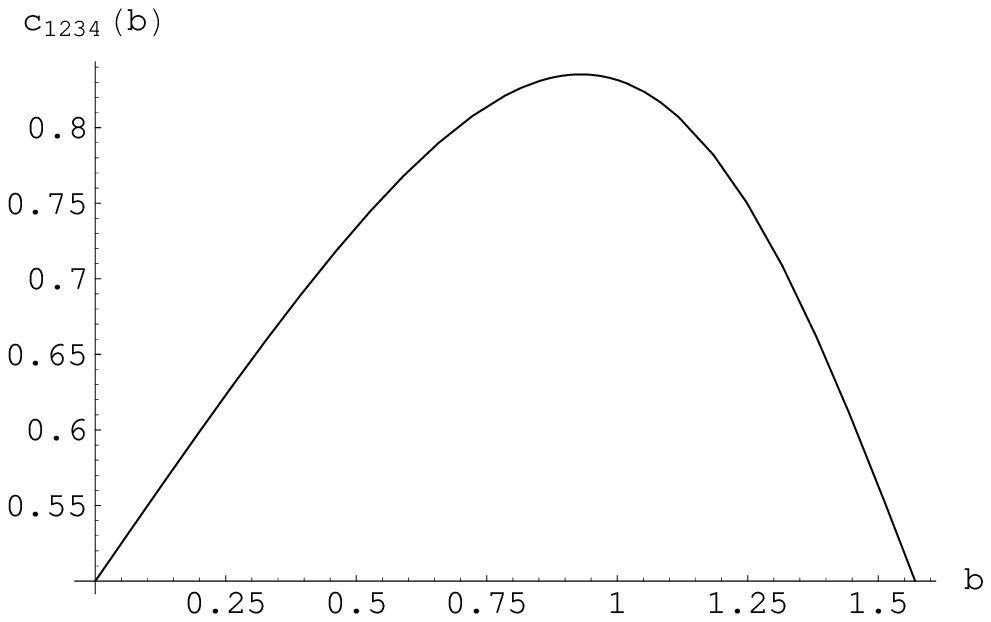}}
\caption{Univariate generalization (\ref{h3}), 
incorporating the Euler angle $b$,
of the constant coefficient (\ref{kdoe}) of $\zeta_{1234}$ in the 
self-dual four-form 
$\Omega_{+}(q_{2})$ (\ref{s1}), corresponding to 
the density matrix $\rho_{2}$
(\ref{denmat2}) }
\label{f3}
\end{figure}
\begin{figure}
\centerline{\psfig{figure=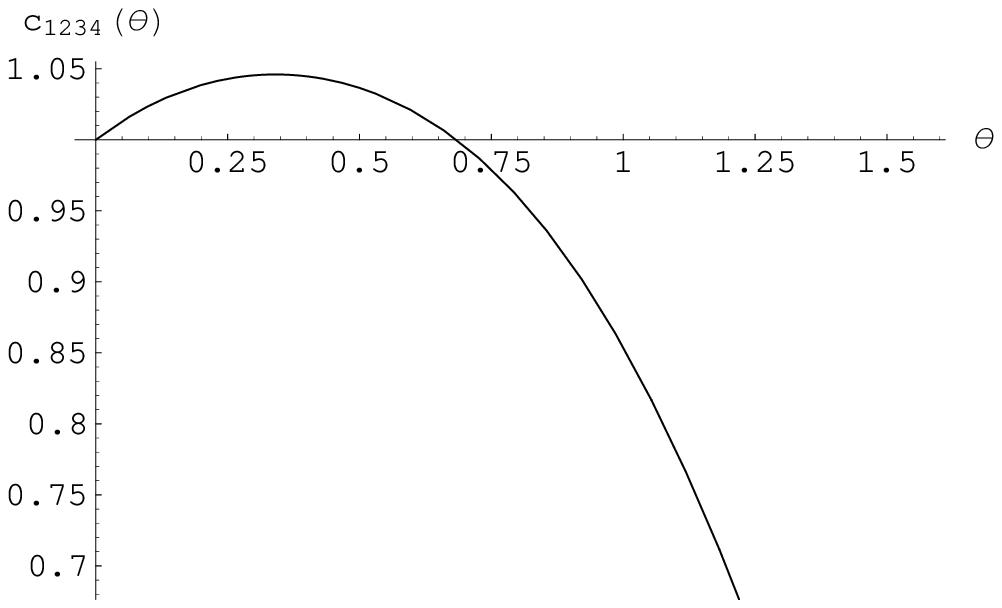}}
\caption{Univariate generalization (\ref{h4}), incorporating the Euler angle
$\theta$, of the constant coefficient (\ref{kdoe}) of $\zeta_{1234}$
in the self-dual four-form $\Omega_{+}(q_{2})$ (\ref{s1}), 
 corresponding to the density matrix
$\rho_{2}$ (\ref{denmat2})}
\label{f4}
\end{figure}
\begin{figure}
\centerline{\psfig{figure=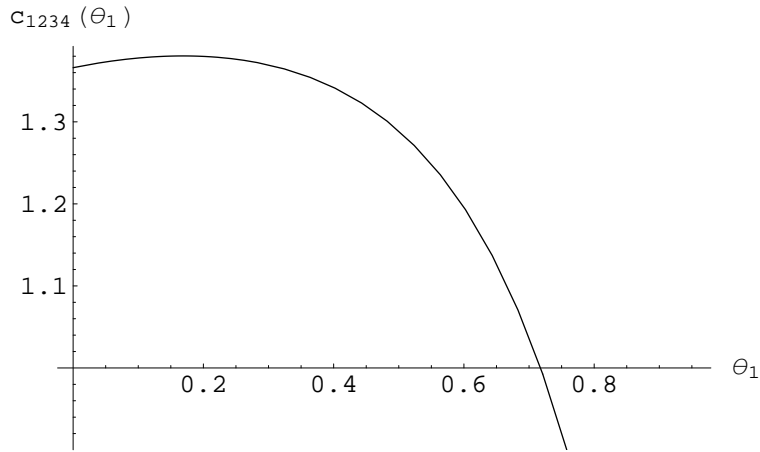}}
\caption{Univariate generalization (\ref{h5}), incorporating the
angle $\theta_{1}$, of the constant coefficient (\ref{kdoe}) of
$\zeta_{1234}$ in the self-dual four-form $\Omega_{+}(q_{2})$ 
(\ref{s1}), corresponding to the density matrix $\rho_{2}$ (\ref{denmat2})}
\label{f5}
\end{figure}
The maximum of Fig.~\ref{f2} is precisely the value (\ref{kdoe}), that is
.821249. 
(We already noted that the twenty-eight eigenvalues associated with the
point $q_{2}$ are unchanged as $\beta$ varies, in contrast with the other
four variables utilized here.)
The maxima for the other four figures are $\tau=.821472,b=.83522,
\theta=1.04598$, and $\theta_{1} = 1.38035$, 
all of which exceed .821249. This last value thus falls 
(as it must) within
the ranges of values assumed by the functions 
above of the form $c_{1234}()$.
(Let us also remark that the specific values of the eight parameters
employed to derive $\rho_{1}$ do not all strictly fall within the 
relatively narrow ranges --- devised so as to avoid duplication --- 
designated for them in \cite{slaterjgp1}. We have so far been unable to
fully determine the corresponding eight parameters, within these designated
ranges, yielding $\rho_{1}$. The point corresponding to $\rho_{2}$, on 
the other hand, does lie within the domain of parameters specified in
\cite{slaterjgp1}.) 

As a futher illustration of the apparent simple characteristics of
the self-dual four-forms in our analytical context,
we give the counterparts of (\ref{kdoe})-(\ref{h5}) and Figs. 1-5, but based
on the 
coefficient of $\zeta_{1678}$ rather than of $\zeta_{1234}$,
\begin{equation} \label{kfoe}
c_{1678} = {-3 + 224 \sqrt{6} \over 6} \approx 90.9476,
\end{equation}
\begin{equation} \label{i1}
c_{1678}(\tau) = 112 \sqrt{{2 \over 3}} +({1 \over 7} + 16 \sqrt{2})
\cos{2 \tau} + \cos{\tau} \sin{\tau},
\end{equation}
\begin{equation} \label{i2}
c_{1678}(\beta) = 
\cos{2 \beta}  + {1 \over 6} (-3 + 224 \sqrt{6}) \sin{2 \beta},
\end{equation}
\begin{equation} \label{i3}
c_{1678}(b) = -{1 \over2} + 56 \sqrt{{2 \over 3}} \csc{b} \sec{b},
\end{equation}
\begin{equation} \label{i4}
c_{1678}(\theta) = -\cos{\theta} + 28 \sqrt{2} \csc{\theta} \sec{\theta}.
\end{equation}
Also,
\begin{equation} \label{i5}
c_{1678}(\theta_{1}) = -{1 \over 2} + {16 (13 \sin{\theta_{1}} + 
\sin{3 \theta_{1}}) \over \sqrt{3} (1 + 7 \cos{2 \theta_{1}})^2}.
\end{equation}
\begin{figure}
\centerline{\psfig{figure=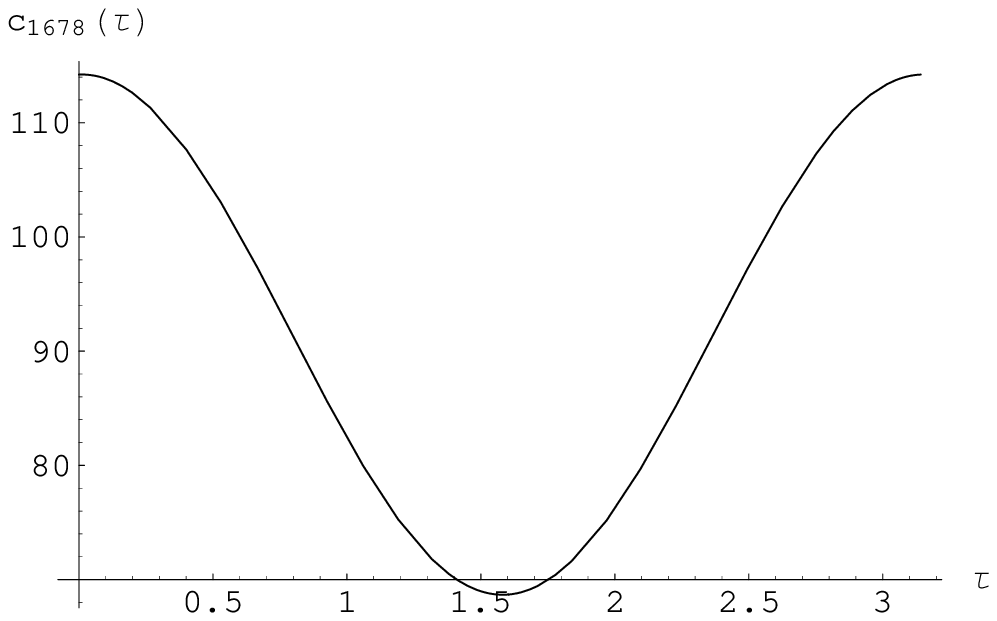}}
\caption{Univariate generalization (\ref{i1}), incorporating the Euler
angle $\tau$, of the constant coefficient (\ref{kfoe}) of 
$\zeta_{1678}$ in the self-dual four-form $\Omega_{+}(q_{2})$
(\ref{s1}), corresponding to the density matrix $\rho_{2}$ (\ref{denmat2})}
\end{figure}
\begin{figure}
\centerline{\psfig{figure=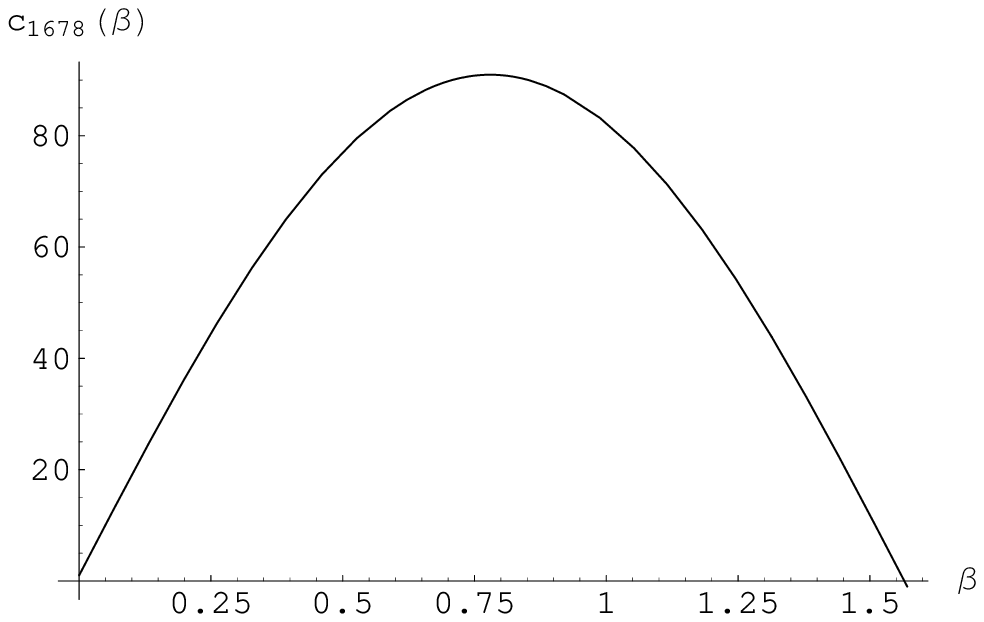}}
\caption{Univariate generalization (\ref{i2}), incorporating the Euler
angle $\beta$, of the constant coefficient (\ref{kfoe}) of 
$\zeta_{1678}$ in the self-dual four-form $\Omega_{+}(q_{2})$
(\ref{s1}), corresponding to the density matrix 
$\rho_{2}$ (\ref{denmat2})} 
\end{figure}
\begin{figure}
\centerline{\psfig{figure=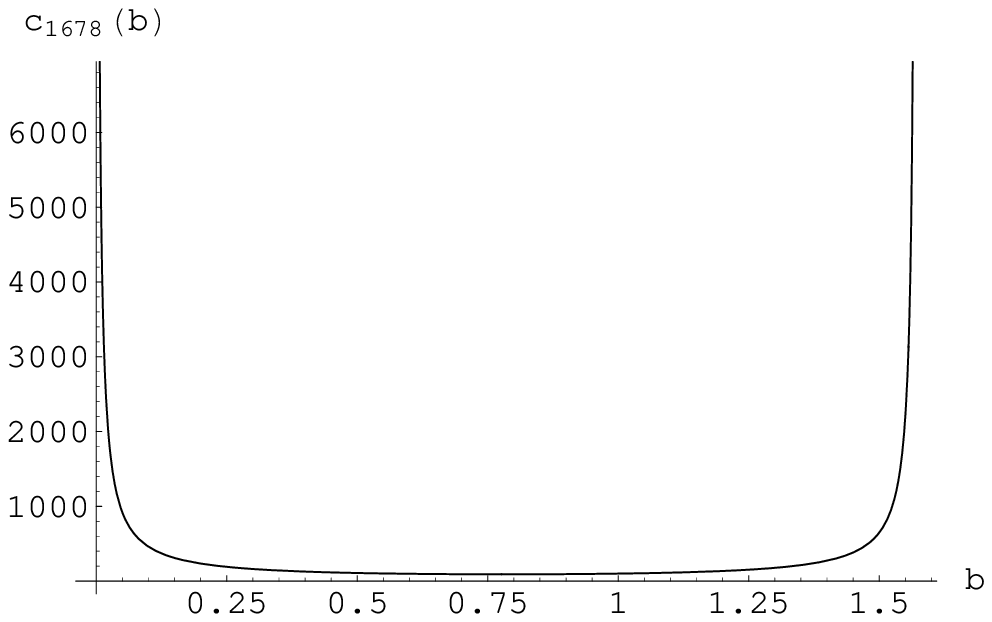}}
\caption{Univariate generalization (\ref{i3}), 
incorporating the Euler angle $b$, of the constant coefficient (\ref{kfoe})
of $\zeta_{1678}$ in the self-dual four-form $\Omega_{+}(q_{2})$
(\ref{s1}), corresponding to the density matrix 
$\rho_{2}$(\ref{denmat2})}
\end{figure}
\begin{figure}
\centerline{\psfig{figure=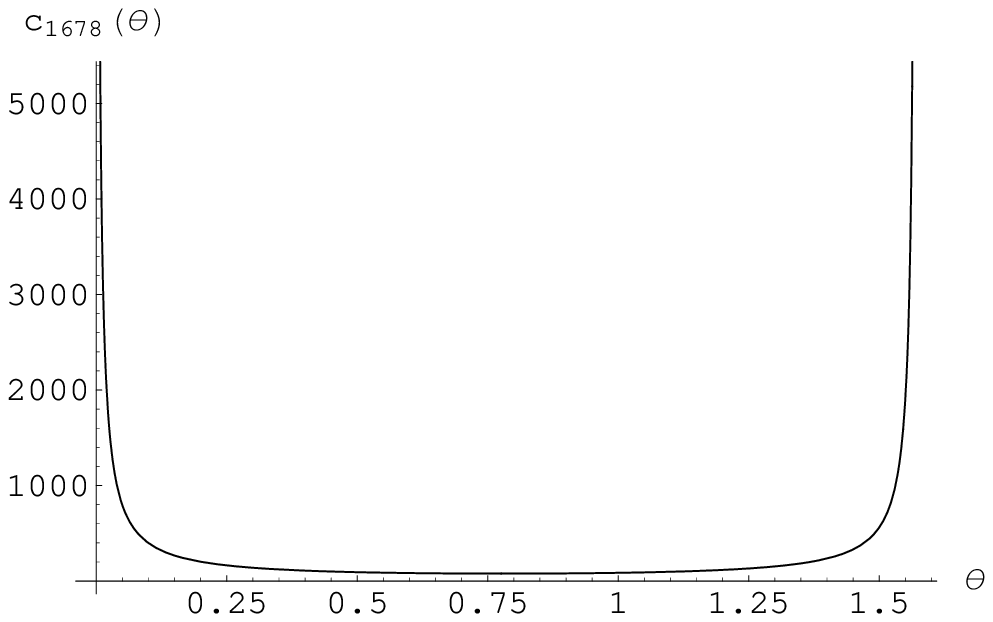}}
\caption{Univariate generalization (\ref{i4}), incorporating the Euler
angle $\theta$, of the constant coefficient (\ref{kfoe}) of
$\zeta_{1678}$ in the self-dual four-form $\Omega_{+}(q_{2})$
(\ref{s1}), corresponding to the density matrix 
$\rho_{2}$ (\ref{denmat2})}
\end{figure}
\begin{figure}
\centerline{\psfig{figure=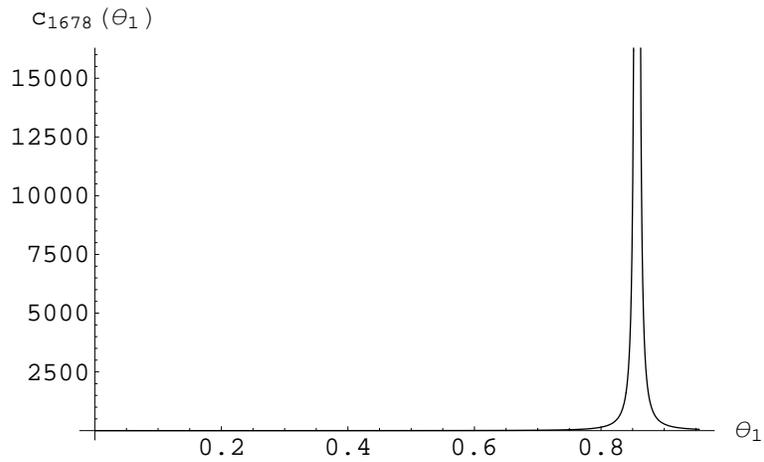}}
\caption{Univariate generalization (\ref{i5}), incorporating the angle
$\theta_{1}$ of the constant coefficient (\ref{kfoe}) of 
$\zeta_{1678}$ in the self-dual four-form $\Omega_{+}(q_{2})$
(\ref{s1}), corresponding to the density matrix 
$\rho_{2}$ (\ref{denmat2})}
\end{figure}

In all this line of work, we hope to further simplify the existing
formulas 
\cite{slaterjgp1} 
for the elements of the Bures metric over the $3 \times 3$ 
density matrix and of its inverse, 
so that the challenging computations of the type reported above 
might be facilitated and further advanced to include the simultaneous
consideration of additional variables.

By way of comparison/contrast of our results, let us note that for the
well-known (self-dual) ``Cayley calibration'' \cite{ff1,corrigan}, 
\begin{equation}
\Upsilon = \zeta_{1234} + \zeta_{1258} -\zeta_{1267} + \zeta_{1357} 
+ \zeta_{1368} - \zeta_{1456} + \zeta_{1478}
\end{equation}
\begin{displaymath}
+\zeta_{2356} -\zeta_{2378} +\zeta_{2457} + \zeta_{2468} -
\zeta_{3458} + \zeta_{3467} + \zeta_{5678},
\end{displaymath}
which is invariant under a 
$\mbox{Spin}_{7}$ subgroup of $\mbox{SO}_{8}$, the $\mathbf{\underline{28}}$
or adjoint representation $\mathfrak{so_{8}}$ of $\mbox{SO}_{8}$
breaks up as
\begin{equation}
\mathbf{\underline{28}} \rightarrow \mathbf{\underline{7}} \oplus  
\mathbf{\underline{21}},
\end{equation}
where the $\mathbf{\underline{21}}$ corresponds to the adjoint representation
$\mathfrak{spin_{7} \in so_{8}}$.
The endomorphism $\hat{\Upsilon}$ of the space of two-forms 
($F$) obeys the characteristic polynomial,
\begin{equation}
\Big( \hat{\Upsilon} - {\mathbb{I}} \Big) \Big(\hat{\Upsilon} + 
3 {\mathbb{I}} \Big) = 0.
\end{equation}
The eigenvalues are, therefore, 1 and -3, and (consistent
with tracelessness)
have multiplicities 21 and 7, respectively. ``Therefore there are two
possible extensions of self-duality, and hence two possible extensions
of the notion of instanton to eight dimensions'' \cite{ff1}.
(The analogous characteristic polynomials of a 
 number of other possible generalized self-dualities --- complex, 
special lagrangian, complex lagrangian, quaternionic, and 
sub-quaternionic --- in
eight-dimensional Euclidean space are also given
in \cite[sec. 6]{ff1}.)
Of course, it would be desirable to develop explanations comparable to
these
of the results we have presented above.

\acknowledgments

I would like to express appreciation to the Institute for Theoretical
Physics for computational support in this research, as well as to
J. M. Figueroa-O'Farrill for his interest and critical insights.

\end{document}